# APPROPRIATION, COLONIALITY, AND DIGITAL TECHNOLOGIES. OBSERVATIONS FROM WITHIN AN AFRICAN PLACE

Gertjan van Stam, Masvingo, Zimbabwe, gertjan@vanstam.net

**Abstract:** This paper provides an assessment of experiences and understanding of digital technologies from within an African place. It provides philosophical reflections upon the introduction and existence – appropriation – of digital technologies. Digital technologies are inherently linked to a colonialising power and, in general, unaligned with local, African ways of *knowing*. Imported technologies are set in modern, universalised doing and unsensitive to the importance of aligned *being* in African contexts. Sensitivities, it is argued, can be fostered by a decolonial turn, where focus shifts from *the individual* to *the community*.

## 1.      PREAMBLE

This paper contributes on the theme of decolonisation in information systems. The paper approaches matters from transdisciplinary (Du Plessis, Sehume, & Martin, 2013) and critical ethnographic angles (Madison, 2012). The argument is built by linking Heidegger's views on *appropriation* to the field of digital technologies and reflect from the author's 20 years of longitudinal observations on digital technologies and academics in African places (van Stam, 2017b, 2019b). Experiences in the natural world and humanities were interrogated and reflected upon following the living research method (van Stam, 2019a). Research, analytics, and dissemination were primed in orality (Mawere & van Stam, 2017; van Stam, 2017a) and communities (Mawere & van Stam, 2020).

This work aligns with a dynamic and integrative epistemology, integrating differentiated experiences, embodied understandings, value judgements, and actions while residing in rural and urban areas in sub-Saharan Africa (Bigirimana, 2017). The work is embedded in a reflexive science where observations, experiences and learnings amalgamate through reflection and introspection, facilitated by theory (Burawoy, 2009). The aim is to recognise patterns and wrestle African understanding out from under a Eurocentric gaze, as acts of decolonisation (Hlabangane, 2018) and epistemic liberation (Buskens & van Reisen, 2016; Grosfoguel, 2011). As such, the work seeks decentred, inclusive, multifaceted understandings, emancipation of polyvocality (the consideration of many voices), diversity and multiple perspectives, from an African positionality (Adamu, 2020).

## 2.      INTRODUCTION

The philosopher and phenomenologist Martin Heidegger (1996), in his seminal work *Being and Time*, uncovers an important field of inquiry when assessing information and communication technology (ICT)-driven innovations, decolonial perspectives, and the experiences of people who hegemonic narratives keep at the margins of ICT development discourses. This field concerns the *appropriation* of being. Heidegger's *appropriation* refers to how existence – *what is* – reveals itself, by itself, without being influenced by any specific categorisation (Mei, 2009).

Reflecting on *how we know what exists* is an important precursor when assessing the role of digital technology in indigenous contexts. It involves uncovering biases in how realities emerge and how they reveal themselves, a field where literature cognisant of African worldviews is scarce (van Stam, 2017b, pp. 40–47). Negating such a reflection signals ignorance, or is an act of *actively forgetting* (Sheehan, 2015). In his efforts to gain an understanding of how reality reveals itself, Heidegger found he could not ignore the issue of *appropriation*. This resulted in, among other things, the emergence of his critical view of technology (Heidegger, 1977). The appropriation of 'mobile' telephony in Africa is well explained through *constitutive appropriation*, which is where systems and artefacts are imported and integrated into the lived experiences of individuals and *communities* (Odumosu, 2018). Here, the word *constitutive* is added to indicate the holistic and embedded nature of the act of appropriation of the technology, for instance, in Africa. Heidegger's critical ontological positions seem to resonate with views on technology that emerge from Africa unframed by Western hegemony (Mavhunga, 2018). Like work done in Australia (Schultz, 2018), African studies reflecting on Heidegger's establishment-challenging views on technology and those emerging from Africa, through conversations and embedded and decolonial thinking, can augment the indigenous journey to digital technologies that are plural and hospitable.

Reflections on the issue of the *appropriation* of technologies and practices is conspicuously absent in digital technologies, especially in situations of export of technologies. In many African cultures, the manner in which a person or action has been *introduced* defines subsequent engagements (du Toit, 2009). *The introduction* positions a *new* entity with an identity in the past and a current and future network of relationships. Therefore, *the introduction* has political implications. The introduction is the start of a continuum of shared being (*having a relationship*). The negation of how digital technologies have been introduced does, therefore, limit gaining understanding of its relational position and the way it is perceived to contribute to the social tapestry in African places. Unfortunately, global arrangements in digital technologies are universally characterised – limited – by widespread 'techno-*think*' and 'techno-*do*' (Sheehan, 2015). Positioning digital technologies as self-explanatory or as a *silver bullet* overlooks the fundamental mismatch of cultural views on its *raison d'être*. The omission of a careful introduction of digital technologies and failing to address the question 'how do we think we know' results in ambiguity on the position of digital technologies in African social structures and among African authorities. This insensitivity links in with a long history of the commodification of African humanity and a disrespect for political authorities and metaphysical perspectives in Africa (Banda, 2019). Therefore, the question of appropriation is crucial when digital developments are contemplated or proposed, as it influences whether or not such interventions can be successfully exported from one philosophical realm to another.

Understanding what digitalisation *means* necessitates assessing how it came about and which authority is vouching for – introducing – its *appropriation*. These are important issues with significant ramifications in digital practice in Africa, as digital technologies are more than mere tools. Digital technologies sustain an environment and structures a way of performing functions in societies (Akrich, 1997). These observations need careful assessment, as the failure to consider digital technologies' social embedding is reminiscent of the introduction of technologies to aid colonial practices.

> *[African p]hilosophy [...] is a critically aware explorative appropriation of our cultural, political, and historical existence.* (Komo, 2017)

> *Philosophy will necessarily have a political dimension. It is naïve to think that philosophy is politically neutral. So, the struggle for liberation becomes the main challenge of every African philosophy which pretends to be hermeneutics. It has to be the hermeneutics of this particular historical situation in Africa – colonialism and neocolonialism: "[…] unlike Heidegger, for us, the question of our existence, of our "to be", is an inherently political question.* (Serequeberhan, 1994)

When considering the cross-cultural setting in which digital technologies and interventions based upon such technologies manifest, Heidegger's abstract philosophical ruminations force the practical question that frames this paper: How are digital technologies affected by coloniality? And, subsequently aiming for usefulness: How can we decolonise digital technologies in Africa? These questions refer to, among other things, the transfer of the cultural consciousness into computer systems and computer human integration. In relation to this, digital services are viewed as a natural extension of the society they serve (Yacob, 2004). In this line of thinking, digital technologies are not just positioned as entities that fill a void (by their sheer existence reifying themselves) but positioned as an agency that aims to augment what is already in place. Of course, in Africa, digital services arrive in already populated spaces of specialists, structures, systems, and practices. Digital technologies *do not liberate* Africa from a condemned past, as per the rhetoric of development, driven by its need for change and innovation, but can be relationally positioned to improve on what is going well.

The African philosopher Achille Mbembe regards the contemporary technologising of the world as a means of spreading Western culture to the rest of the world (Mbembe, 2017). In his paper *Bodies as Borders* (Mbembe, 2019), he describes a *computable speed regime*, where society is organised around the forces and energy derived from *computability*. This self-perpetuating regime thrives on normative descriptions in a mathematical language, which are, in turn, transposed into algorithms for derivations. He links this focus on quantifiable information and governance through calculations to how technology is used to set up mobile, portable, omnipresent and ubiquitous delineations and borders. In this sense, technology functions as the means by which *the West manages the Rest* (Adamu, 2021; Mamdani, 2012). Mbembe's observations of people being controlled through technology are not new. The Americans Paul Dourish and Scott Mainwaring (2012) point out how the thoughts underpinning ubiquitous computing are inherently colonial. They argue that this thinking metastasises technologies to act redemptively based on the following assumptions:

- The need to 'civilise': Centres of power (hubs, labs) are geared towards assisting places perceived as 'lacking'.
- Superiority: Development is undertaken by people/entities that are assumed to be knowledgeable or powerful.
- Universality: What works *here* can be applied *there*, with same results.
- Reductive truth: Comparison, evaluation, understanding and prediction follow from quantification and statistics.
- Future making: Centres of power/*the developed* are the blueprint for progress in other regions, for *the developing*.

The risk in these assumptions is that those who assume that they possess power define what is *true* in terms of the thinking by those who are subdued (Mamdani, 2012). Such external truths cause a clash of paradigms, also in digital technologies (Mawere & van Stam, 2015).

Technologies carry expressed values, from the site of their conceptualisation to the place of deployment (Feenberg, 2005). This imposition of culture through technology is easy to understand when assessing drones or killer robots, but less obvious when implementing digital apparatus like mobile phones and Wi-Fi routers, or interacting with applications that exist by virtue of the dashboard. It is no wonder that the Cape Townian Shaun Pather and his Latin American co-author Ricardo Gomez (2012), in an evaluation of years of publications on information and communication technologies, were prompted to wonder: 'Do we ask the right questions?'

Current philosophies of science and the resulting methodologies and mainstream academic practices seem not to recognise, or uniformly negate, inputs from African conceptions (Mawere & van Stam, 2015). Dominant science and technology theories are predominantly Western (Harding, 1998). From this exogenous gaze, in digital technologies, African communities are often approached as non-agentic spectators. In such a view, of course, scientists and foreign specialists feel warranted to develop and implement solutions that they deem best. Although *doing good* is set in method and attitudes of benevolence, its judgement is set by intellectuals and professionals operating in frames derived from a knowledge system foreign to Africa. This knowledge solidifies in methods and policies on how one *should* think about the materialisation of *the digital* and use in African contexts. This orientalistic system obscures the fact that judgement of *what is best* is set by dominant, imperial powers in an Eurocentric thought positioned to explain *others* (van Stam, 2020). Such a patronising *Eurosplaning* does little, if anything, to align with analysis and judgement from African knowing and viewpoints (Adamu, 2019). The resulting digital *solutions* can only be sustained by (en)forced attention. Prime examples of forced attention are workshops in which attendants are being paid to attend and the setting of agendas on development, training, measurement and evaluation according to priorities and measures of success enforced from the outside.

African communities do harbour appropriate knowledge, embedded in and from the context. This knowledge outlasts any contribution by someone from the outside, who comes and is prone to leave at any time (soon). In contrast, when digitalisation emerges from within it is sustained through ownership by local communities (Bidwell, 2020).

African knowing is not only subalternised, it is generally not recognised, unwanted and seldom asked for (Jeater, 2018). When reviewing publications by non-African scientists, one hardly finds reference to the thinking and contributions of Africans. However, in academic works produced in Africa, one can see many references to non-Africans and explicit reference to contributions set in non-African philosophies. The disregard of African philosophies and practices by the dominant philosophies of science is discomforting when residing in Africa. It represents an epistemicide of African meaning making (Grosfoguel, 2010, 2013; Mawere, 2014; Nyamnjoh, 2012). Scholars such as Mexican Enrique Dussel (Dussel, 2000) and South African Mogobe Bernard Ramose (2016), among others, call for the resurrection of local ways of knowing and respect for the underlying worldviews. Such an orientation can withstand the uncritical import of Eurocentric paradigms in Africa. Mind boggling questions arise, such as: How can we decolonise digital technologies based on philosophies that have hardly, if any, African input? And, even when recognising this chicken-and-egg problem, how can we provide decolonised digital technologies in general?

## 3. AN AFRICAN NARRATIVE

The definition of *colonialism* is, like most definitions, set by the powerful. Although there are numerous definitions, the common denominator is that they are all proposed from the perspective of the coloniser. When assessing colonialism from the perspective of the colonised, it looks quite different. Colonialism, in this perspective, is a circular scheme of shaming, brainwashing, and conditionality, in other words: derogative judgment, demands for mimicry, and shackling (van Stam, 2017c). In Africa, contemporary dominating practices manifest as *super-colonialism* – a scaled-up version of old-style colonialism, perpetrated not only by countries sustaining a capitalist elite, but more so by corporations and international organisations acting as if they were sovereign (Mbembe, 2019). The practices of super-colonialism can be seen in the workings of the global monetary system, international trade treaties, and the monopolisation of cyberspace. All of these systems strive to keep Africa subjugated and dancing to its tune (Mamdani, 2012). From this perspective, the importing of digital technologies and interventions from outside Africa is set in the continuation of the colonial narrative of *bringing civilisation* and *modernity* to a presumably *dark* and *backwards* Africa.

In addressing the issue of decolonisation in higher education, South Africa has been a torch bearer. Under the banner of the #RHODESMUSTFALL and #FEESMUSTFALL movements, students in South Africa have demanded decolonisation in higher education (Nyamnjoh, 2016). Although norms seem to be changing, it remains challenging to talk about conquests, the common praxis of ontological and epistemic violence, and, ultimately, an imperial academy with a long history of supporting colonial acts. The significant benefits resulting from the continuing extraction of Africa's resources, including data, provides no incentive to recognise the colonially-inspired structure of contemporary information systems and their architectures (Mawere & van Stam, 2016a). A comparison between the old colonial shipping routes and the architecture of the Internet reveals an incredible similarity. The architecture of the Internet resembles the global information society as it was created in the 16th and 17th centuries, in which information was collected in African and brought in ships to knowledge centres in the West, where it was analysed and used (Mawere et al., 2019). Contemporary Internet infrastructure follows the shipping routes that colonial operatives used to extract this information.

Some describe the act of decolonisation as *doing the right thing* – whatever that may be. This is a questionable concept, as the causes of *doing* are set in character and how one conducts oneself. In turn, character is set in histories of experience, understanding, judgements and action. Addressing domination needs inspection, retrospection and introspection. Therefore, decolonisation requires an accurate and honest assessment of power, privilege, history and the present. Ethics dictate that, regarding digital technologies, we must:

*[…] think through what it might mean to design and build computing systems with and for those situated at the peripheries of the world system, informed by the epistemologies located at such sites, with a view to undermining the asymmetry of local-global power relationships and effecting the 'decentering' of Eurocentric/West-centric universals.* (Ali, 2016)

This is not a task for the fainthearted. In fact, it can put one's position in the (colonised) society on the line. There is a lot of resistance to positions that challenge dominant standpoints and the status-quo. There is no easy fix for social injustices like racism, discrimination, and (digital) inequality. Unfortunately, the process of judging often reifies the problem.

## 4.     THE ACADEMY

Academia, especially in Africa, appears to thrive on *copy and paste* practices. For instance, the philosophies and methodologies valorised through hegemonic and universalising lenses are often uncritically transplanted and regurgitated as *good science* in Africa. The contemporary practices of parachute researchers crowd-out the work of academics in their local context (Bockarie, 2019; Mawere & van Stam, 2019).

The hegemonic, Eurocentric academy is regarded as a veritable institution – part of the system of good governance of knowledge, also in Africa. And yet, this academy has consistently failed to live up to this promise, as it invariably reinforces dominant narratives about a Eurocentric modernity (Koch & Weingart, 2016). As a result, universalised science does not hold true in most African contexts. In practice, it often serves as a vehicle for conquest and control (Mawere & van Stam, 2019). Subsequently, the academy is not understood by the majority of people in the so-called *periphery*. Being discarded to the margins, there are few ways of holding an imperial academic elite accountable for not respecting African ways of meaning-making.

Although there is much diversity in academy, it appears that many academics fear the label *activist*. They adhere to the dichotomy of theory and practice. Clearly, theoretical outcomes result in the rearranging of power in practice: technologies configure who can do what and where. This makes technology (e.g., cryptography and artificial intelligence, which are important technical components) political tools (Rogaway, 2015). Therefore, science must consider *appropriation* and the moral dimensions of its exploits (Murphy & Ellis, 1996). Unfortunately, Eurocentric structuring underpins *ivory tower* academics, a situation that is rarely questioned in Africa (Mawere & van Stam, 2019). A significant part of the academy believes that science is inherently good and *value independent*, although indications to the contrary are mounting. Apparently, many academics are unaware of the long and guilty history of upholding *the Terrible Three*: an othering *orientalism*, hegemonic *imperialism* and occupying *colonialism* (van Stam, 2017b), often in the name of the mythical concept of *modernity* (Dussel, 2000).

African study centres in Europe started off supporting colonial conquests. However, this history is overlooked in their external communications; apologies for their support for the *colonial project* have not been forthcoming. The brainwashing of "hegemonic education can exist only so long as true and accurate information is withheld" (Asante, 1991, p. 177). The battle for hegemony takes place in African education; here the adoption and regurgitation of Eurocentric, soulless, calculable reasoning and knowledge is being primed, because "…the curriculum [is] the terrain of the struggle for the transformation of the educational paradigm" (Ramose, 2003, p. 140). It is clear that decolonisation is relevant in every nook and cranny of society, and thus also in the digital realms, as its technologies replicates its underlying models of thinking.

## 5. TECHNOLOGY AND DECOLONIALITY

Most power structures, including those in academia, are the continuation of the colonial matrix of power, set in hegemonic epistemologies and systems (Ndlovu-Gatsheni, 2013). When assessing the powers-that-be from a colonial perspective, it appears that colonialism never ended (van Stam, 2017c). Africans are continually being shamed, brainwashed and usurped by hegemonic, foreign-inspired socio-economic narratives and structures. In addition, Africans are increasingly the subject of imported techno-social systems under the guise of securitisation. Digital platforms seek to analyse every African word and breathe to fuel the apparatus of their super-colonial masters. Corporate entities seeking *to know it all* are operating as de facto sovereignties and transnational institutes, or colluding with security forces from *the centre* (Greenwald, 2014). The means of appropriating resources are being perfected in a super-colonialism enabled by ubiquitous connectivity that surpasses the colonialism that relied upon the Westphalian concept of a nation state.

> *While it is important to challenge Eurocentric epistemologies in texts, it is not enough to do so if structural and physical factors of the colonial world help create and maintain the same epistemology that scholars are currently trying to decolonise.* (Richardson, 2018)

Inherited from the colonial ethnology and maintained by ignorance and lack of interest, prejudices about Africa still run high (D'Almeida-Topor, 2006). The structure of many digital developments builds upon a one-sided view of economics, democracy – or information and computing for development for that matter – that emerged from imperialistic practices (Ndlovu-Gatsheni, 2013). Academic practice appears rather amnesic to its coloniality (Grosfoguel, 2011). The seeking of African understandings of the meaning and framing of the topic *digital technologies in Africa* – in language, philosophy and worldview – is subdued and, when proposed, subalternised (Louai, 2012). This discordant situation is fuelled by a lack of comprehension of – and even contempt for – indigenous perspectives and narratives, and a general lack to seek such a *comprehension* (Arendt, 1968; Mawere & van Stam, 2017). Consequently, for Africa, in general, the prospective benefits from foreign research and development and the resulting technical assistance processes are problematic (Mawere & van Stam, 2019).

From his study of Frantz Fanon as a psychiatrist who wrote about psychology, Nelson Maldonado-Torres (2017) argues that a *decolonial turn* puts attitude over method. This results in a transdisciplinary practice that mixes well with decolonial activism. Current digital practices, however, position method before attitude. Attitude is soft power in action, gained from an accumulation of experiences, understandings, and judgements. Attitude is an aspect of dynamic, integrative meaning making (Bigirimana, 2017). Such a meaning making involves the ongoing accumulation of insights and interlinks action with values and ethics. The content of these components varies per location. It is different in Africa than in other areas of world (van Stam, 2012). Therefore, African understanding of digital technologies will be different to the understanding of digital technologies elsewhere.

A decolonial turn is fundamental and all encompassing. It represents a paradigm switch, not a paradigm shift (van Stam, 2017b). In Africa, paradigm switching allows an oriental *projected self,* linked with universalism, to be replaced by a locally-grounded African *authentic self,* linked with diversity. The oriental self is composed of normative statements that measure, classify, and datafy African lives. This essentialising (re)territorialises and creates social silos and spatial identities, as well as relationships between space and bodies on the basis of abstract universals. In contrast, the African authentic self is based on value epistemologies like *ubuntu* or *ifa*, being the crystallisation of African philosophies and communal love (Mawere & van Stam, 2016b). African identities are embedded in, for lack of better words, indigenous wisdom traditions and local, regional and diverse narratives of (trans)modernity.

In the I-Paradigm, the *individual* is the unit of analysis. An individual is a human being independent of social groups or relationships. This, of course, is a theoretical concept, as each person is intrinsically connected, having emerged into the realm of the living through a woman's womb and existing in an interdependent web of relationships. In this definition, the proto-*individual* is a human being that views him/herself (and others view him/her) independently from his/her relationships – a specific (and solitary) self in a world consisting of many other distinct (and solitary) selves (van Stam, 2021). This understanding of individualism aligns with assessments of how Western culture and so-called modernity "calls for the limiting of oneself in one's private, egotistical 'me', with a tightly isolated circle where one can satisfy one's own urges and consumer whims" (Kapuściński, 2008). *Individualism* and universality seem to be the dominant paradigms that guide contemporary renderings of digital technologies. Universality is claimed by a normative epistemology, in which knowledge is positioned as certain, indubitable, infallible and incorrigible, as well as objective. This knowledge is acquired by a *knower* who is a passive and neutral spectator of a detached reality. As a result, such knowledge is proposed as an accurate representation of the *real* world, ready to be boiled down into formulas for computation.

In Africa, the single human being is understood as *a person*, a human entity within a social network of relationships with companions, ancestors, the living, and the yet to be born, as well as non-humans. Initiation confirms that the person understands what it means to be *in union* with society and the more extensive social system – what one's part and position is within it. Grounded in those understandings, one's rights can be exercised, and duties performed in connection and set in systems of moral behaviour according to the espoused cultural values. A person continually evaluates and adjusts their personal conduct to the known or assumed expectations of those with whom he or she relates. In this way, the person shifts the focus of their conduct from an individualised-self to a communal-self, contributing to the development and upkeep of shared values. A community is a gathering of persons who subscribe to shared set of values; it provides belonging, is the embodiment of values and beliefs, is ethically grounded, and exists in a confined area in which justice can be sought (van Stam, 2021).

The We-Paradigm – the paradigm of *interbeing* – sees universalised, normative knowledge as external to the community, and is quick to link such knowledge to external belief systems, power and domination. Dynamic, integrative knowledge, however, is what the community *knows* in connection (Mawere & van Stam, 2017; van Stam, 2013). Exogenous digital technologies appear to have great difficulty recognising such a paradigm, in which community is paramount, where *knowing* is an accumulation of insights, gained through the active involvement of the knower by experiencing, understanding, judging and acting, involving interactions within the whole organisms, not in isolated minds – and in which creative expressions like art, dance, ritual, and sculpture are considered among the highest forms of knowledge.

Universalised views on digital technologies continue to be proposed for deployment in contemporary situations in Africa. As they are conceptualised in line with interests positioned outside of Africa, this hegemony serves colonial interests. Decoloniality is, thus, a proper topic for contemplation in relation to digital technologies in Africa. Colonialism exists because of concepts such as individualism, the concept of a social contract, and the capitalist economic system. Upon switching the paradigm of perception from *individualism* (the I-Paradigm) to *community* (the We-Paradigm) – with its social-personhood –issues of colonialism fall away. The We-paradigm is incompatible with notions of supremacy, hegemony, and domination. It is grounded in notions of progress, movement and the raising of consciousness, connected with emotions, senses and embodied experiences. The communal paradigm is, therefore, incompatible with colonialism. Switching paradigms in digital technologies is, thus, a redemptive act of decolonisation. It is an ideological switch from digital technologies set in self-proclaimed universal truths to a digitisation embedded in context, cognisant of communities and diversity, conviviality, grace, and frontierness. This type of digitalisation transcends a focus on individual benefits and the concepts that give rise to colonialism to put morality, human beings, values, listening and feelings in their rightful place – at the locus of meaning-making and subjective insights (du Toit, 2007), the centre of the synthesisation of the rational and imaginative, the immeasurable and quantifiable, and the intellectual and emotional.

A paradigm *switch* to community sets digital developments on a pathway towards diversity to foster unity and synthesis, transcendence (in time and place), and the reconciliation of seemingly incommensurable views. It results in digital technologies that contribute to a contextually-aligned understanding of what life encompasses. This involves the letting go of a digitisation that is solely seen through a universal lens and based on Eurocentric valorisation, in a self-consciousness that regards authenticity and the individual (and, therefore, individualism) as the primary agent interacting in communities. Decolonisation addresses issues of method, aligning the method of research and development with the geographical context and manner in which knowledge is pursued in that place. In Africa, methods incorporate the management of relationships (especially in situations of conflict), and the harmoniousness of relationships between humans, nature, and the divine (Louis, 2007).

Just as the idea that a universalised view of digital technology needs to be challenged, likewise, the appropriation of data and displacement of information from Africa to non-African centres of computing must be recognised for what they are: colonial practices (Harris, 1998). These notions are gaining traction in post-colonial studies in the humanities (Smith, 1999), however, they are not readily accepted in the natural sciences.

Digitalisation and its technologies sit at the intersection of the natural sciences and the humanities. The consideration of social and environmental justice is crucial in a world that appears to be headed towards ecological destruction. A decolonised digital realm values ethnocentricity and the reciting of stories that constitute being; it seeks to allow the *appropriation* that Heidegger described. For instance, artificial intelligence, considered a promising aspect of digital technologies, is a misnomer, as, for instance, machines have no notion of mortality (Olivier, 2017). Prioritising the social aspects of digital technologies allows *knowing* to emerge and assesses experience from ethical standpoints – it values the journey. Above all, it questions the function of (imported) methods and theories in African places. It acknowledges that information has many facets that need to be assessed in the local context for it to be inclusively productive and integrally helpful (van Stam, 2013). The illusion of universalising science – which is a location-dependent (European), historically-determined, contingent ethno-science – suffocates the voices, visions, and contributions of a great variety of other genuine knowledge systems and epistemologies from the many cultures in Africa. Acknowledging that all knowledge is local knowledge decolonises and takes the *sting* out of modernity's pretence at monolithic trajectories for digital technologies.

African ways of knowing and gaining knowledge are important, and the world needs them (Metz, 2020; Salami, 2020). For instance, an African way of seeking consensus was used during the Durban Conference on Climate Change in 2011. During tough times, under enormous pressure, the climate conference in Paris in 2015 was pulled out of the muck using *indabas*, which is a type of conference or consultation practised in South Africa (Rathi, 2015). Although the strength of the accord is yet to be seen, it is clear that without the introduction of this African conversation technique, there would never have been an accord at all.

Some say that communal practices have been lost, communities are gone, and uniformity is the norm. Such a demise might well be witnessed in many urbanised worlds. However, as mentioned before, in Africa, most people do not live in urban areas. Long-established communities have been cut by colonial borders, imported notions of nation-state, and migration (Otieno, 2018). This destruction of communities is being replicated in cyberspace. In African environments, the sense of community is certainly not lost. This puts a decolonised digital narrative at odds with hegemonic, universalised views.

# 6. CONCLUSION

In Africa, digital technologies cannot be governed by systems and methods of intervention that Africans have had no part in conceptualising. Over centuries, Africa's precious contributions and vital experiences have been neglected in a Eurocentric academy and its off springs from their research and development. Such indigenous contributions and experiences are known within, and from, a communal paradigm. This has deprived digital developments of precious and important information. Paradigm switching replaces colonially-framed knowledge with something that is already in place: communal understandings of reality. Switching the paradigm from *I* to *We* is necessary in movements to decolonise digital technologies. Such a decolonised digital realm builds upon the strength of sharing, care, respect, and maturity.